\documentclass[aps,pre,twocolumn,showpacs,amssymb,amsmath,superscriptaddress]{revtex4-1}
\usepackage{graphicx}
\usepackage{txfonts}
\usepackage{aurical}
\usepackage{color}
\usepackage{soul} 
\graphicspath{{./figures/}}
\DeclareGraphicsExtensions{.pdf}
\newcommand{\figwidth}{\linewidth}
\newcommand{\dif}{\mathrm{d}}

\def\coola{\mbox{\Fontauri A}}
\usepackage{comment}

\definecolor{myblue}{rgb}{0,0,1}

\usepackage[pdftex, colorlinks=true, linkcolor=myblue, citecolor=myblue,
urlcolor=myblue]{hyperref}

\newcommand{\ipcms}{Universit{\'e} de Strasbourg, CNRS, 
Institut de Physique et Chimie des Mat{\'e}riaux de Strasbourg, UMR 7504, F-67000 Strasbourg, France}

\newcommand{\csic}{Instituto de Estructura de la Materia, IEM-CSIC, Serrano 123, Madrid 28006, Spain}

\begin{document}

\title{Correlation between peak-height modulation and phase-lapses in transport through quantum dots}

\author{Rodolfo A.\ Jalabert}
\affiliation{\ipcms}
\author{Rafael A.\ Molina}
\affiliation{\csic}
\author{Guillaume Weick}
\affiliation{\ipcms}
\author{Dietmar Weinmann}
\affiliation{\ipcms}


\begin{abstract}
We show that two intriguing features of mesoscopic transport, namely the modulation of Coulomb blockade peak-heights and the transmission phase-lapses occurring between subsequent peaks, are closely related.
Our analytic arguments are corroborated by numerical simulations for chaotic ballistic quantum dots. 
The correlations between the two properties are experimentally testable.
The statistical distribution of the partial-width amplitude, at the heart of the previous relationship, is determined, and its characteristic parameters are estimated from simple models.  
\end{abstract}



\maketitle

\section{Introduction}

The low-temperature electronic transport through a sub-micron quantum dot (QD) is dominated by two effects. Firstly, the Coulomb repulsion between electrons on the dot leads to an energy cost for adding an extra electron to the QD. Due to this charging energy, the tunneling of electrons to or from the reservoirs can be Coulomb blockaded \cite{kastner92,kouwenhoven97,hanson07}. Secondly, the confinement in all three directions leads to quantum effects that sign the conductance \cite{imry,alhassid,scholarpedia}. The importance of these two effects greatly depends on the size of the QD (i.e., the number of electrons that can be varied from one to several thousands), as well as the openness of the QD (i.e., the coupling to the reservoirs that can be weakened by the effect of tunnel barriers). 
The small and weakly connected QDs operating in the regime of Coulomb blockade (CB) present fascinating physical properties that can be exploited for diverse functionalities, like the manipulation of the spin in few-electron systems \cite{hanson07} or low-temperature thermometry \cite{hahtela17}.

Despite being an effect due to electron-electron interactions, CB allows in many cases a simple description where the interaction between electrons in the dot and those in the environment are parametrized by a constant capacitance which does not depend on the electron filling \cite{kouwenhoven97}. Such an approach, called the constant interaction model (CIM), results in a one-particle problem with a constant charging energy that electrons must ``pay'' for entering the QD. 

Throughout the CIM, the properties of one-particle wave-functions within the QD become relevant for the description of quantum transport in the CB regime. Such a connection, together with the assumption that the underlying classical dynamics of electrons in the dot is chaotic, or at least sufficiently far away from regular, has been exploited by proposing the view of QDs in the CB regime as laboratories for studying quantum chaos issues \cite{scholarpedia}.

The wave-function distributions and correlations in chaotic systems is a subject of sustained interest in the context of quantum chaos studies \cite{berry77,backer08}. In particular, the one-particle wave-function distribution has been shown to be a key ingredient of two emblematic problems of mesoscopic physics: the CB peak-height fluctuations and the transmission phase of a QD. In the first case, the distribution of CB peak-heights has been related with the universal statistical properties of chaotic wave-functions \cite{jala92}. Experiments performed using QDs defined in two-dimensional electron gases have validated such a relation \cite{chang_96,folk_96}, but the observation of a long-energy-range modulation of the peak-height distribution indicated the existence of departures from the universal behavior \cite{hackenbroich97,vallejos99,narimanov}.    
In the second case, the phase-lapses of the QD transmission amplitude have been related with the signs of the wave-function at the entrance and the exit of the QD \cite{levy95,lee99,taniguchi99,gefen99,levy00,silvestrov00,hacken}.
Experimentally, the conductance of Aharonov-Bohm rings was measured, with a QD embedded in one of the arms. The phase-lapses in the energy-dependent transmission amplitude of a QD were inferred from the observation of a phase-locking of the Aharonov-Bohm flux dependence of the ring conductance when the QD was tuned to consecutive CB peaks \cite{jacoby95,schuster97,avinum05}. 

The statistical properties of the wave-function distribution have been shown to result in a phase-locking of the conductance that is not perfect, but more and more likely to be observed as the semiclassical limit of large $k L$ is attained \cite{molina12,molina13,jalabert14}. Here,
$k$ is the electron wave-vector in the QD and $L$ is the distance between its entry and exit (see Fig.\ \ref{fig:setup}). 

While the early experiments \cite{jacoby95, schuster97} showed sequences of perfectly universal phase evolution with phase-lapses in every conductance valley (except for extremely low electron fillings in the dot \cite{avinum05}), recent measurements using a new way to reliably extract the transmission phase from the conductance of an Aharonov-Bohm interferometer \cite{takada} detected several conductance minima without phase-lapses \cite{takada17,edlbauer2017}. Though it has not been possible to achieve a complete statistics in these last experiments, the findings are consistent with the predicted departure from the universal behavior of a perfect phase-locking of conductance peaks~\cite{molina12,jalabert14}. 

In this work we consider the link between the two above-described and seemingly unrelated problems, showing a correlation between the modulation of the peak-height distribution and the absence of phase-lapses in the transmission amplitude through a QD. The use of the CIM allows us to derive such a connection from the statistical properties of the one-particle wave-functions and of the partial-width amplitudes (PWAs) that quantify the coupling between the QD wave-functions and the propagating waves in the leads. 

Our paper is organized as follows: The formalisms for the peak-height and the transmission amplitude, together with the conjecture that relates the two, are presented in Sec.~\ref{sec:transmissionphase}. Section \ref{sec:correlation_numerics} provides the numerical verification of the correlation between the modulation of peak-heights and the absence of phase-lapses. The statistical distribution of the PWAs, at the heart of the verified correlation, is determined from numerical simulations in Sec.~\ref{sec:pwa_numerics}, where we also discuss the pertinence of its description by a Gaussian distribution. In Sec.~\ref{sec:bc_pwa_correlator} we use a simple model system to calculate the parameters that characterize the PWA distribution. We conclude in Sec.\ \ref{sec:ccl}.
In the Appendix we go beyond the case of a chaotic QD and provide results for an integrable geometry, the rectangular billiard.

\section{Transmission phase and peak-heights in the Coulomb blockade regime}
\label{sec:transmissionphase}

\begin{figure}[tb]
\centerline{\includegraphics[width=0.6\linewidth]{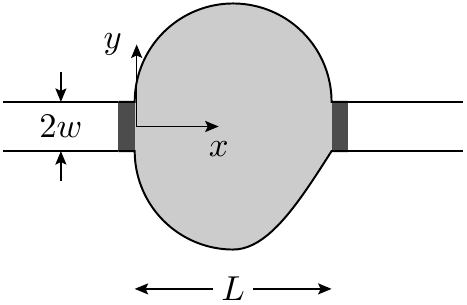}}
\caption{\label{fig:setup} Sketch of an asymmetric two-dimensional quantum dot (light gray) connected to leads through tunnel barriers (dark gray), used in the numerical calculations. $L$ is the separation between entrance and exit of the dot and $2w$ is the width of the leads.}
\end{figure}

Within an effective one-particle description, enabled by the adoption of the CIM, the R-matrix theory can be used to relate the transmission through the QD with the eigenstates of the closed system \cite{jala92,alhassid}. In the weak-coupling regime, where the average resonance width $\Gamma$ is much smaller than the mean level spacing $\Delta$ of the QD, the transmission amplitude $t$ for an incoming energy $E$ is dominated by Breit-Wigner contributions corresponding to the resonances that are nearest in energy, and reads 
\begin{equation}
\label{eq:Breit-Wigner}
t(E) = \sum_{n} \frac{\gamma_n^{\mathrm{l}} \gamma_n^{\mathrm{r}}}{E - E_n + \mathrm{i} \, \Gamma_n/2} \, .
\end{equation} 
For the $n$\textsuperscript{th} level, we denote $E_n$ the resonance energy, $\gamma_n^{\mathrm{l}(\mathrm{r})}$ the PWA (or effective coupling) for tunneling between the QD and the left (right) lead, 
$\Gamma_n^{\mathrm{l}(\mathrm{r})}=|\gamma_n^{\mathrm{l}(\mathrm{r})}|^2$
the corresponding partial width, and $\Gamma_n = \Gamma_n^{\mathrm{l}}+\Gamma_n^{\mathrm{r}}$ the total width of the resonance.
The PWA for tunneling into the left (right) lead is related to the QD wave-function $\psi_n$ of the corresponding resonance by \cite{alhassid}
\begin{equation}\label{pwa}
\gamma_n^{\mathrm{l}(\mathrm{r})} = 
\sqrt{\frac{\hbar^2k_\mathrm{L}P_\mathrm{c}}{M}}
\int_{-w}^{+w} {\rm d}y \ \psi_n(x^{\mathrm{l}(\mathrm{r})},y) \ \Phi_{\mathrm{l}(\mathrm{r})}(y) \, ,
\end{equation}
where $\Phi_{\mathrm{l}(\mathrm{r})}$ is the first transverse subband
wave-function in the left (right) lead (assumed both of width $2w$). The integration is along 
the transverse coordinate $y$ at the entrance (exit) of the QD located at $x=x^{\mathrm{l}(\mathrm{r})}$ (see Fig.\ \ref{fig:setup}), $P_\mathrm{c}$ is the transparency of the tunnel barriers, $k_\mathrm{L}$ is the Fermi wave-vector in the leads, and $M$ is the electron effective mass. We assume the absence of magnetic field, so that $\psi_n(x,y)$ and the PWAs can be taken to be real.

At zero temperature, the Landauer-B\"uttiker formula \cite{imry} and the Breit-Wigner transmission amplitude \eqref{eq:Breit-Wigner} lead to a Lorentzian form for the energy-dependent conductance $G(E)$ close to the $n$\textsuperscript{th} resonance. For each spin channel, we thus have
\begin{equation}\label{ztc}
G(E)=\frac{G_0}{2} \ \sum_{n} \frac{\Gamma_n^{\mathrm{l}} \ \Gamma_n^{\mathrm{r}}}
{\left(E - E_n\right)^2+\left(\Gamma_n/2\right)^2}\, ,
\end{equation}
where $G_0=2e^2/h$ is the quantum of conductance.

For the $n$\textsuperscript{th} resonance, the zero-temperature peak-height can be characterized by the dimensionless quantity 
\begin{equation}\label{PH0}
g_n(T\!=\!0) = 
\frac{G^\mathrm{peak}_n(E_n)}{G_0/2} \simeq 
\frac{4 \Gamma_n^{\mathrm{l}} \ \Gamma_n^{\mathrm{r}}}
{\Gamma_n^2}\, .
\end{equation}
For finite temperatures $T$, such that $\Gamma \ll k_\mathrm{B}T \ll  \Delta$, the dependence of the conductance on the Fermi energy in the leads is given by the derivative of the Fermi distribution, and the dimensionless peak-height is given by \cite{beenakker_91}
\begin{equation}\label{PHT}
g_n(T)=\frac{\pi}{2k_\mathrm{B}T} \ \alpha_n 
\, ,
\end{equation}
where we have defined
\begin{equation}\label{alpha}
\alpha_n  = \frac{\Gamma_n^{\mathrm{l}} \ \Gamma_n^{\mathrm{r}}}
{\Gamma_n}\, .
\end{equation}

The wide variation of peak-heights experimentally observed at low temperatures among sequences of CB peaks \cite{meriav_90,kouwenhoven_91} called for a proper statistical ensemble where the peak-height distribution can be studied. Since the CB physics yields well-separated peaks in energy, 
it has been proposed~\cite{jala92} to generate the statistical ensemble by considering a given peak number $n$ within a collection of irregular billiards with the same area $\mathcal{A}$. The statistics of $\alpha_n$, and then that of $g_n(T)$, are determined by the fluctuations of the partial widths $\Gamma_n^{\mathrm{l}(\mathrm{r})}$ with respect to their average value $\langle \Gamma_n/2 \rangle$ over equal-area irregular billards. Similar considerations hold for $g_n(T\!=\!0)$ given in Eq.~\eqref{PH0}. The situation of symmetric couplings between the dot and the two leads, which yields the same average value of the two partial widths $\langle \Gamma_n^{\mathrm{l}}\rangle = \langle\Gamma_n^{\mathrm{r}}\rangle$, is adopted for simplicity. 

The average level-width $\langle \Gamma_n \rangle$ has a tendency to increase with $n$, since the tunnel barriers become more transparent as the energy of the incoming electrons increases. Such monotonic secular behavior as a function of $E_n$ can coexist with modulations on energy-scales which are much larger than $\Delta$. Experiments \cite{folk_96}, semiclassical approximations \cite{narimanov}, and numerical simulations \cite{jalabert14} have put in evidence these variations. We tackle this important issue in the next section.

Switching to the problem of the phase evolution between the $n$\textsuperscript{th} and $(n+1)$\textsuperscript{st} CB peaks, we distinguish the case where the transmission amplitude at intermediate energies is dominated by the contributions arising from these two resonances, from that with large fluctuations of the level width among different levels, where far-away resonances might characterize the intermediate behavior. The first scenario supposes the absence of strong fluctuations of the level-widths, and has been termed \cite{molina13} restricted off-resonance (ROR) behavior. The second scenario of large PWA fluctuations, termed unrestricted off-resonance (UOR) behavior, is expected (and numerically verified \cite{molina13,jalabert14}) to be rare.

In the energy evolution of the transmission amplitude between two resonances, each passage through the origin of the complex plane (zero of the transmission) is associated with a phase-slip of $\pi$. Within the ROR behavior, the transmission amplitude vanishes between the $n$\textsuperscript{th} and $(n+1)$\textsuperscript{st} resonances if, and only if, the sign rule \cite{levy00}  
\begin{equation}
\label{eq:condition}
D_n=\gamma_n^{\mathrm{l}}\gamma_n^{\mathrm{r}}\gamma_{n+1}^{\mathrm{l}}\gamma_{n+1}^{\mathrm{r}} >0
\end{equation}
is verified. 
In the case of the UOR behavior, the previous rule implies an odd number of transmission zeros between two resonances, which results in the phase-locking of the conductance between adjacent peaks. Therefore, the difference between ROR and UOR behavior is not relevant concerning the phase-locking phenomenon. 
The absence of a phase lapse is related with a non-vanishing of the transmission between two resonances.
Recently, such a behavior has been experimentally verified for relatively large quantum dots \cite{edlbauer2017} through the observation 
of missing phase-lapses correlated with non-negligible values of the conductance between the adjacent resonances.

The condition \eqref{eq:condition} depends on the correlation of the signs of the PWAs associated with adjacent peaks  \cite{footnote:nodal_lines}. The likelihood for the absence of phase-slips between resonances can be characterized by the probability $\mathcal{P}(D_n<0)$ of finding $D_n<0$.

Since we consider billiards with irregular geometries and Hilbert spaces with large dimensionalities, we neglect possible eigenvector (resonance wave-function)
correlations. Therefore, denoting ${\cal P}(\gamma_n^{\mathrm{l}} \gamma_n^{\mathrm{r}} > 0)$ the probability of having the two PWAs of the $n$\textsuperscript{th} resonance with the same sign, we have
\begin{align}
\label{eq:probDn}
{\cal P}(D_n<0) =&\; 
{\cal P}(\gamma_n^{\mathrm{l}} \gamma_n^{\mathrm{r}} > 0)\big[1 - {\cal P}(\gamma_{n+1}^{\mathrm{l}}
\gamma_{n+1}^{\mathrm{r}} > 0)\big]
\nonumber \\
&+
{\cal P}(\gamma_{n+1}^{\mathrm{l}} \gamma_{n+1}^{\mathrm{r}} > 0)\big[1 - {\cal P}(\gamma_{n}^{\mathrm{l}}
\gamma_{n}^{\mathrm{r}} > 0)\big] \ .
\end{align}

The statistics of the peak-heights is determined by the partial-width distribution, while the condition $\mathcal{P}(D_n<0)$ results from the PWA distribution. The two above-cited distributions are obviously linked, and therefore the peak-height statistics and phase-lapses must also be related to each other. 

Independently of the precise statistical distribution of the wave-functions, in view of Eq.\ \eqref{pwa} it seems reasonable to assume that the first moments of the PWA distribution vanish, i.e., 
$\langle \gamma_n^{\mathrm{l}(\mathrm{r})} \rangle = 0$. 
Within the previous assumption of neglecting correlations of wave-functions corresponding to different resonances, 
we have 
$\langle \gamma_n^{\mathrm{l}(\mathrm{r})} \gamma_{n'}^{\mathrm{l}(\mathrm{r})} \rangle = 0$ for $n \neq n'$. 
For each $n$, the first non-vanishing moments of the PWA distribution are the second-order ones 
$\langle \gamma_n^{\mathrm{l}} \gamma_n^{\mathrm{l}} \rangle$, $\langle
\gamma_n^{\mathrm{r}} \gamma_n^{\mathrm{r}} \rangle$, and $\langle
\gamma_n^{\mathrm{l}} \gamma_n^{\mathrm{r}} \rangle$. Under the hypothesis of symmetric couplings between the leads and the QD, these second moments are characterized by the two parameters
\begin{equation} 
\label{eq:sigma_def}
\sigma_n^2 = \langle \gamma_n^{\mathrm{l}}\gamma_n^{\mathrm{l}} \rangle =
\langle \gamma_n^{\mathrm{r}}\gamma_n^{\mathrm{r}} \rangle 
= \langle \Gamma_n/2 \rangle
\end{equation}
and
\begin{equation} \label{eq:rho}
\rho_n = \frac{1}{\sigma_n^2} \
\langle \gamma_n^{\mathrm{l}}\gamma_n^{\mathrm{r}} \rangle \ .
\end{equation}

Assuming that the average values have a smooth dependence on $n$, the products $\gamma_n^{\mathrm{l}}\gamma_n^{\mathrm{r}}$ and $\gamma_{n+1}^{\mathrm{l}}\gamma_{n+1}^{\mathrm{r}}$ are likely to have the same sign provided that 
$\langle \gamma_n^{\mathrm{l}}\gamma_n^{\mathrm{r}} \rangle = \rho_n \sigma_n^2$ is not close to zero. The absence of phase-lapses, related to the occurrence of $D_n<0$, is more likely to be obtained when $\langle \Gamma_n \rangle$ and/or $\rho_n \sigma_n^2$ are small. A small  $\langle \Gamma_n \rangle$ results, through Eqs.\ \eqref{PH0} and \eqref{alpha}, in low conductance peak-heights $g_n(T\!=\!0)$ and $g_n(T)$.
\textit{We therefore conjecture that the absence of phase-lapses in the transmission phase between two conductance peaks is correlated with a low value of the average height of these peaks.} 

\section{Correlation between the peak-height modulation and the absence of phase-lapses: numerical results}
\label{sec:correlation_numerics}

In this section we perform numerical simulations to demonstrate the connection between the peak-height modulation and the absence of phase-lapses conjectured above. The model system and the numerical calculations follow those of Ref.~\cite{jalabert14} for spinless non-interacting electrons. 

In order to gather a good statistical sampling with well-defined average values, the ensemble of equivalent irregular billiards previously defined can be complemented by a spectral average over a collection of consecutive resonances. In this case, care has to be applied concerning the energy-dependence of the mean values. The overall increase of $\langle \Gamma_n \rangle$ with $E_n$, together with the modulations on energy scales much larger than $\Delta$ discussed in Sec.~\ref{sec:transmissionphase}, limit the spectral average to a narrow window $\Delta \varepsilon$ where the mean values are weakly energy-dependent.  

In the ensemble of equivalent irregular billiards completed with local spectral averages, the weakly energy-dependent mean total width is defined by
\begin{equation} \label{eq:mtw}
\Gamma_{\varepsilon} = \frac{1}{\varrho_{\rm sc}(\varepsilon)} \ 
\sum_{n} \langle \Gamma_n \ 
\delta_{\Delta \varepsilon}(\varepsilon - E_n) \rangle \  .
\end{equation}
As before, the brackets represent the average within the collection of equivalent irregular billiards. The function $\delta_{\Delta \varepsilon}$ is a normalized window for selecting the resonance energies. One possible choice is 
\begin{equation} \label{eq:enwin}
\delta_{\Delta \varepsilon}(\xi) = 
\frac{1}{2 \Delta \varepsilon} \ 
\Theta \left(\Delta \varepsilon - |\xi| \right)
 \  ,
\end{equation}
where $\Theta$ is the Heaviside step function (normalized Lorentzian or Gaussian functions are also appropriate \cite{toscano}), while 
\begin{equation} \label{eq:dos}
\varrho_{\rm sc}(\varepsilon) = - \frac{1}{\pi} \ 
{\int_\mathcal{A} \ \mathrm{d} \mathbf{r} \ {\rm Im} \ G_{\rm sc}(\mathbf{r}, \mathbf{r}; \varepsilon)}
\end{equation}
is the smooth (Thomas-Fermi) part of the density of states in the QD, 
with $G_{\rm sc}$ the small-$\hbar$ limit of the Green function.
Analogously to the expression \eqref{eq:mtw} for the mean total width $\Gamma_{\varepsilon}$, we can define the local moments of the PWA distribution, and from them, the energy-dependent parameters $D_{\varepsilon}$, $g_{\varepsilon}(T\!=\!0)$, $g_{\varepsilon}(T)$, etc. 

Experimentally, the extended ensemble is obtained by following the CB peaks within a narrow interval of gate voltage under a continuously changing magnetic field (that must remain small enough not to change the symmetry class of the dot wave-functions) and varying plunger gate voltages that modify the geometry of the QD while keeping its area approximately constant \cite{folk_96}. 

For the numerical simulations we consider the asymmetrical stadium billiard of Fig.\ \ref{fig:setup}, where one of the quarter circles is replaced by a cosine curve. 
Changing the potential inside the billiard at fixed Fermi energy, as it could be done by the effect of a back-gate voltage in an experiment, allows to sweep over the resonances of the dot. Fitting the energy-dependence of the numerically-obtained transmission amplitude $t(E)$ with the form of Eq.\ \eqref{eq:Breit-Wigner} allows to extract the energies $E_n$, the corresponding products $\gamma_n^{\mathrm{l}} \gamma_n^{\mathrm{r}}$, and the total widths $\Gamma_n$ for a sequence of resonances in some energy interval. 
The assumption of an unchanged billiard shape under variations of a side- (plunger-)gate voltage has been challenged for experimentally-realizable QDs \cite{hackenbroich97,vallejos99}.

The collection of equivalent irregular billiards is obtained by considering fourteen asymmetric shapes resulting from the deformation of different quarter-circles of the stadium. In the extended ensemble, the resonance energies are labeled by the variable $\varepsilon$ and the corresponding wave-vector $k$, related by $\varepsilon=\hbar^2 k^2/2M$. Instead of an $\varepsilon$-window as in Eq.\ \eqref{eq:enwin}, we adopted a $k$-window and defined the local spectral averages working with $k$-dependent parameters $\Gamma_k$, $D_k$, $g_k(T\!=\!0)$, $g_k(T)$, etc. For a given $k$, the collection of nearby peaks within an interval of length $\pi/4L$ was included in the ensemble. Such $\Delta k$ was chosen as to include many resonances, but being much smaller than $\pi/L$, which is the expected scale of variation for the $k$-depending averages. Based on this expectation, and in order to improve the statistics, the data of six successive $k$-intervals differing by $\pi/L$ were mixed \cite{jalabert14}. An overall energy-shift is performed in each sweep in order to maximize the overlap of the long-range $k$-modulations of the mean values.

\begin{figure}
\centerline{\includegraphics[width=\figwidth]{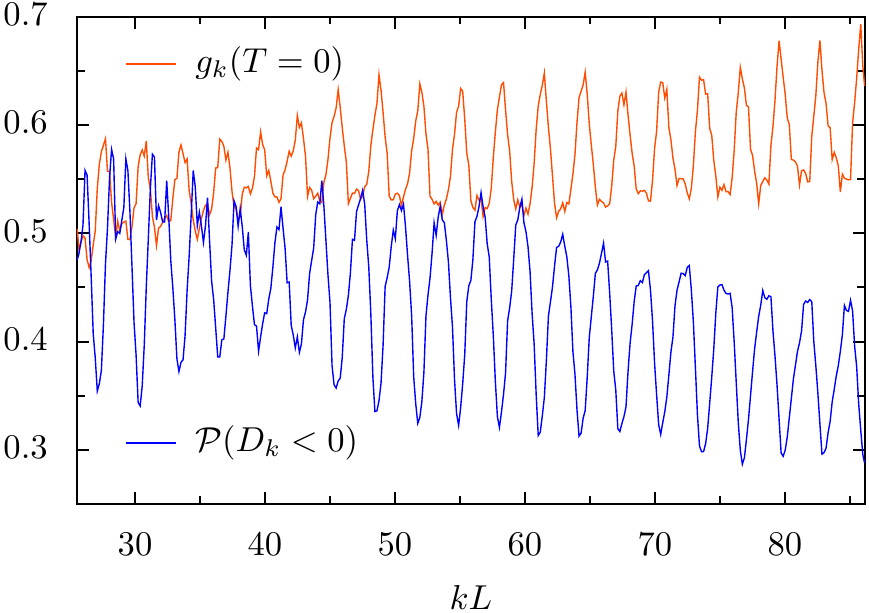}}
\caption{\label{fig:correlation} Probability ${\cal P}(D_k<0)$ of 
having a negative $D_k$ (blue line) and average conductance peak-height (orange line) as a function of $kL$. The most-likely values of $kL$ to observe the lack of phase-locking in the conductance, defined by the condition $D_k<0$, are those where the peak-heights are minimal.}
\end{figure}  

The blue line in Fig.~\ref{fig:correlation} reproduces the probability ${\cal P}(D_k<0)$ as a function of the wave-vector $k$ obtained in Ref.\ \cite{jalabert14} (up to the correction of a non-crucial numerical mistake). The orange line represents the mean peak-height at zero temperature $g_k(T\!=\!0)$ obtained by using the same averaging procedure defined above. Superimposed to the secular increase with the electron energy, $g_k(T\!=\!0)$ exhibits a modulation in the scale of $kL=\pi$. Such a modulation is clearly anti-correlated with the behavior of $\mathcal{P}(D_k<0)$. Maxima of ${\cal P}(D_k<0)$ (absence of phase-lapse) occur when the peak-height is minimal. 

We therefore numerically confirm the conjecture of Sec.\ \ref{sec:transmissionphase} relating the evolution of the transmission phase with the modulation of the resonances' peak-height for the case of chaotic cavities. A similar anti-correlation is found if we consider the average finite-temperature peak-height $g_k(T)$ instead of the zero-temperature one (not shown). 

\section{Statistical distribution of the partial-width amplitudes}
\label{sec:pwa_numerics}

The connection between the peak-height modulation and the absence of phase-lapses discussed in the previous sections is of statistical nature. The joint probability density of the PWAs appears as the key ingredient in both problems. Therefore, in this section we concentrate ourselves in the characterization of such a distribution. A crucial question at this time is whether or not the joint distribution of the PWAs is a Gaussian one. And if this is the case, a second relevant question is that of obtaining the parameters defining such a distribution. 

\begin{figure}
\centerline{\includegraphics[width=\figwidth]{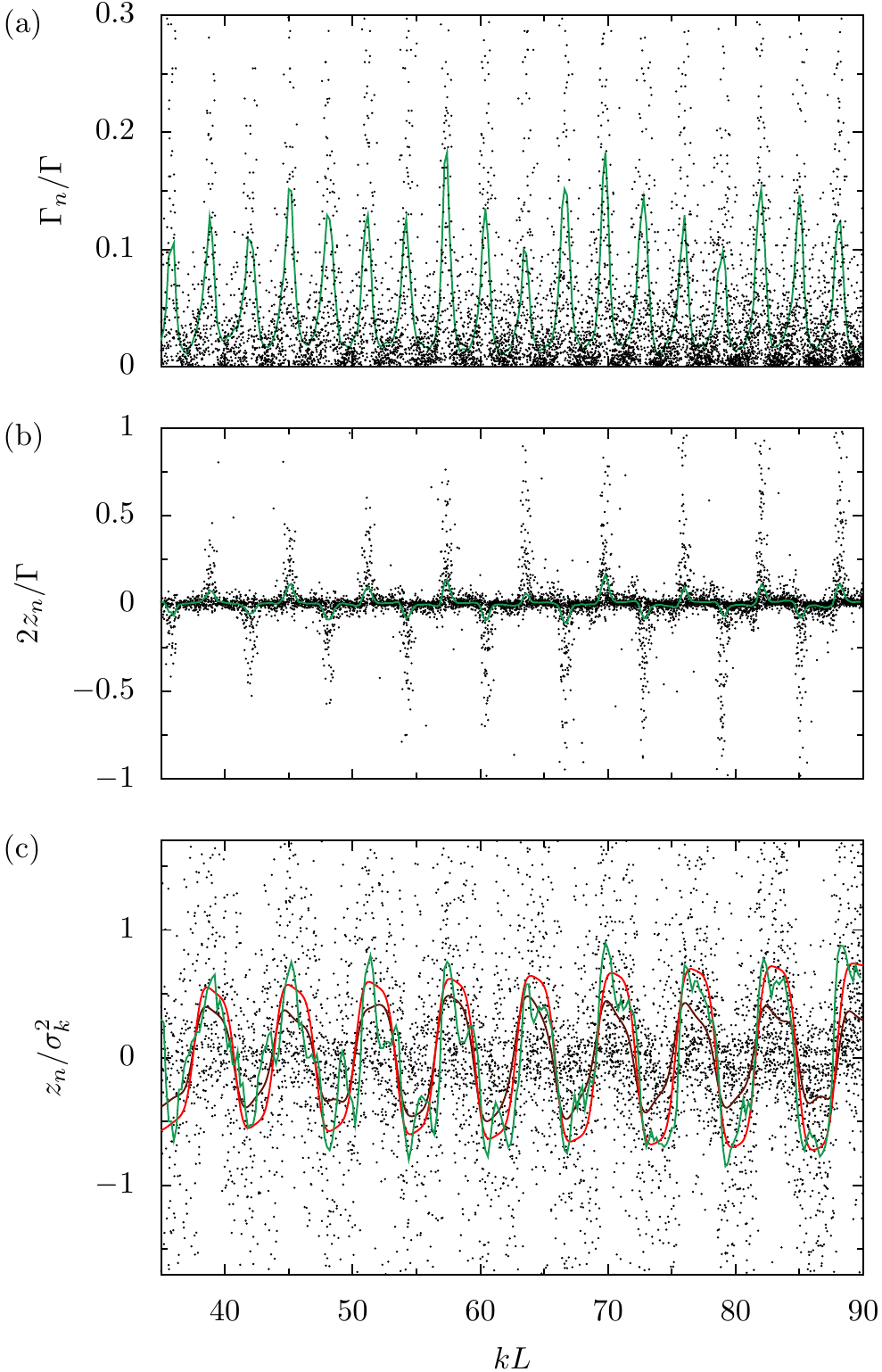}}
\caption{\label{fig:correlator_rho_glr} (a) Individual values of the total width $\Gamma_n=|\gamma_n^{\mathrm{l}}|^2+|\gamma_n^{\mathrm{r}}|^2$ (black dots) of the resonances, characterized by the values $k_nL$ indicated on the horizontal scale,
in units of the global average $\Gamma$ of the total width.
(b) Individual values of $z_n=\gamma_n^{\mathrm{l}}\gamma_n^{\mathrm{r}}$ (black dots), normalized by $\Gamma/2$.
(c) Same data points as in panel (b), but normalized by the local variance $\sigma_k^2$ (black dots). 
The green solid lines in panels (a), (b), and (c) represent, respectively, the values of $2\sigma_k^2$, $2\sigma_k^2\rho_k/\Gamma$, and $\rho_k$ 
obtained by fitting the local $z_n$ distribution by Eq.\ \eqref{eq:probdistrproduct}. In panel (c), the 
correlator $\rho_k$ obtained in Sec.~\ref{sec:bc_pwa_correlator} for the rectangular billiard sketched in Fig.~\ref{fig:rectangle} is shown according to the mean-field approximation \eqref{eq:rhoMFA} (brown solid line) and to the numerical evaluation of Eq.~\eqref{eq:rhofull} (red solid line).}
\end{figure}  
 
The data points in Fig.\ \ref{fig:correlator_rho_glr} (black dots) represent the values of $\Gamma_n$ and $\gamma_n^{\mathrm{l}} \gamma_n^{\mathrm{r}}$, corresponding to different resonance wave-vectors $k$, obtained from the fitting of the numerically-calculated $t(E)$, 
which were used to build the results of Fig.~\ref{fig:correlation}. In Fig.~\ref{fig:correlator_rho_glr}(a), the distribution of 
$\Gamma_n = |\gamma_n^{\mathrm{l}}|^2 + |\gamma_n^{\mathrm{r}}|^2$ is shown in units of the overall average total width $\Gamma$. The distribution of the product 
$z_n=\gamma_n^{\mathrm{l}} \gamma_n^{\mathrm{r}}$ is presented with two different normalizations: using  $\Gamma/2$ [Fig.\ \ref{fig:correlator_rho_glr}(b)] and using the local variance $\sigma_k^2$ [Fig.\ \ref{fig:correlator_rho_glr}(c)] arising from the procedure described below. 

\begin{figure*}
\centerline{\includegraphics[width=.8\linewidth]{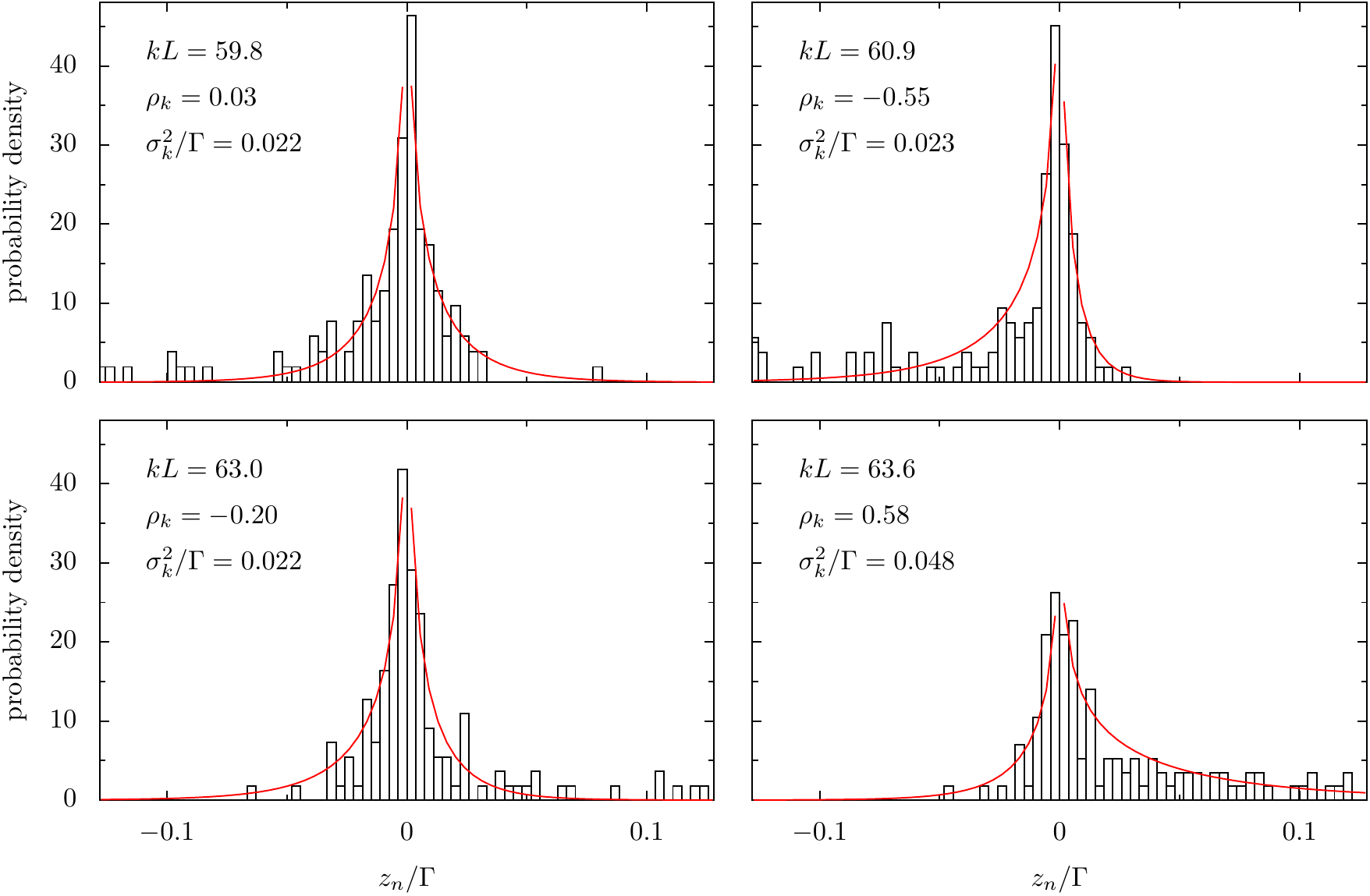}}
\caption{\label{fig:product} Numerically-obtained histograms describing the probability distribution of the product of partial-width amplitudes $z_n$, for four arbitrarily chosen values of $kL$. The red lines represent fits of the numerical data using the expression 
\eqref{eq:probdistrproduct} based on a Gaussian distribution of the PWAs.
The overall average total width $\Gamma$ has been chosen in order to present $z_n$ and $\sigma_k^2$ in dimensionless form.}
\end{figure*}  

The above-discussed modulation in $kL$ is clearly visible in the $\Gamma_n$ and $z_n$ distributions, on the scale of $\pi$ and $2\pi$, respectively.
These distributions have then to be studied for a given $n$, or at most for a $k$-interval much smaller than $\pi/L$. We use the complemented
ensemble to build histograms describing the probability density of $z_n$ \cite{footnote:poorer_stat}. 

The four $k$-values selected in Fig.\ \ref{fig:product} illustrate the strong $k$-modulation of the $z_n$ distribution. In order to characterize the $\Gamma_n$- and $z_n$-distributions, a
Gaussian law for the joint probability density of the left and right PWAs appears as the simplest hypothesis to test. Such a distribution 
is solely characterized by the parameters $\sigma_n^2$ and $\rho_n$ 
defined in Eqs.\ \eqref{eq:sigma_def} and \eqref{eq:rho}, respectively, and reads
\begin{align}
\label{eq:jointprobdistr}
p(\gamma_n^{\mathrm{l}}, \gamma_n^{\mathrm{r}}) =&\,
\frac{1}{2 \pi \sigma_n^2 \sqrt{1 - \rho_n^2}} 
 \nonumber \\
&\times
 \exp{\left( - \frac{\left(\gamma_n^{\mathrm{l}}\right)^2+
\left(\gamma_n^{\mathrm{r}}\right)^2 - 2 \rho_n \gamma_n^{\mathrm{l}}
\gamma_n^{\mathrm{r}}}{2 \sigma_n^2 (1 - \rho_n^2)} \right)} \, .
\end{align}
According to \eqref{eq:jointprobdistr}, the $z_n$ probability density is given by \cite{springer}
\begin{equation}
\label{eq:probdistrproduct}
p(z_n) =
\frac{1}{\pi \sigma_n^2\sqrt{1 - \rho_n^2}} \  
 \exp{\left( \frac{\rho_n \ z_n}{\sigma_n^2 (1 - \rho_n^2)}\right)} \
 K_0\left(\frac{|z_n|}{\sigma_n^2 (1 - \rho_n^2)}\right) \, ,
\end{equation}
where $K_0$ is the zeroth modified Bessel function of the second kind. 

The histograms describing the distribution of the variable $z_n$ of Fig.\ \ref{fig:product} are rather well described by the distribution \eqref{eq:probdistrproduct} arising from a Gaussian law for the joint distribution of the PWAs (red solid lines). The limited sampling that we are able to numerically gather does not allow to settle the 
statistical significance of the departures from Eq.\ \eqref{eq:probdistrproduct} observed, in some cases, for large values of $|z_n|$. 

Fitting the $k$-dependent histograms to Eq.\ \eqref{eq:probdistrproduct} allows to extract local values of $\sigma_k^2$ and $\rho_k$. Figure \ref{fig:correlator_rho_glr}(a)
presents $2\sigma_k^2$ (green solid line) which provides a good description of the typical values of the $\Gamma_n$-distribution and exhibits a large (order of magnitude) modulation in the $k$-variable with a characteristic scale of $\pi/L$. The average values of the $\Gamma_n$-distribution (not shown) are dominated by 
extreme realizations, and thus provide a much poorer description of the data than the value $2\sigma_k^2$ obtained by fitting the histograms. 
In a similar way, the values of $2\sigma_k^2\rho_k/\Gamma$ and $\rho_k$ presented, respectively, by the green solid lines in Figs.\ \ref{fig:correlator_rho_glr}(b) and 
\ref{fig:correlator_rho_glr}(c) exhibit a $k$-modulation in the scale of $2\pi/L$ and provide a good description of the typical values of the $z_n$-distribution when using 
the global and local normalizations. The increase of the typical values of $z_n$ observed in Fig.\ \ref{fig:correlator_rho_glr}(b) upon increasing $k$ is a consequence of the increased transparency of the barriers discussed in Sec.\ \ref{sec:transmissionphase}, together with the use of the global normalization factor $\Gamma/2$.

Given the relatively good accuracy of the Gaussian law describing the joint probability density of the PWAs, it is important to study the underlying hypothesis sustaining such a law and the consequences for the observable  distributions. Moreover, it is important
to determine the parameters characterizing Eq.\  \eqref{eq:jointprobdistr}, 
which could be compared with those stemming from the fitting procedure previously used. In the rest of this section we discuss the founding principles needed to obtain a Gaussian joint probability density of the PWAs and recall the implications of such a distribution for the problem of the transmission phase. The estimation of the correlator parameter is performed in the next section. 

According to Eq.\ \eqref{pwa}, the statistical properties of the PWAs follow from the wave-function distribution for the eigenstates of the QD. The latter can be addressed by employing different approaches. Among them, the maximum entropy principle \cite{jarzynski97} establishes that, given the limited knowledge that an eigenstate exists at energy $\varepsilon$, the least biased guess for the wave-function $\psi_{\varepsilon}$ with a 
two-point correlation function 
$\langle \psi_{\varepsilon}^{*}(\mathbf{r}) \ \psi_{\varepsilon}(\mathbf{r}^\prime) \rangle = {\cal C}(\mathbf{r}, \mathbf{r}^\prime; \varepsilon)$ that respects the normalization condition 
$\int_\mathcal{A} \mathrm{d} \mathbf{r} \
|\psi_{\varepsilon}(\mathbf{r})|^2 = 1$ 
is the Gaussian distribution
\begin{equation}
\label{eq:probdistrwf}
p(\psi_{\varepsilon}) = \mathcal{N} \exp{ \left( -\frac{\beta}{2} 
\int_\mathcal{A} \!\! \mathrm{d} \mathbf{r}
\int_\mathcal{A} \! \! \mathrm{d} \mathbf{r}^\prime 
\psi_{\varepsilon}^{*}(\mathbf{r}) 
\ {\cal K}(\mathbf{r}, \mathbf{r}^\prime; \varepsilon) \ 
\psi_{\varepsilon}(\mathbf{r}^\prime)
\right)} \, ,
\end{equation}
where $\mathcal{N}$ is a normalization constant [independent of $\psi_{\varepsilon}(\mathbf{r})$]. For the case of a system with time-reversal invariance that we treat, we have $\beta=1$ and $\psi_{\varepsilon}^{*}(\mathbf{r})=\psi_{\varepsilon}(\mathbf{r})$. The kernel function ${\cal K}(\mathbf{r}, \mathbf{r}^\prime; \varepsilon)$ is the functional inverse of the two-point correlation function, i.e.,
\begin{equation}
\label{eq:kernel_tpf}
\int_\mathcal{A} \!\! \mathrm{d} \mathbf{r}^{\prime\prime}
\ {\cal K}(\mathbf{r}, \mathbf{r}^{\prime\prime}; \varepsilon) \ 
{\cal C}(\mathbf{r}^{\prime\prime}, \mathbf{r}^{\prime}; \varepsilon) = 
\delta(\mathbf{r}-\mathbf{r}^{\prime})
 \, .
\end{equation}
Equations \eqref{pwa} and \eqref{eq:probdistrwf} lead to the Gaussian law \eqref{eq:jointprobdistr} for the joint probability density of the PWAs. The parameters $\sigma_n^2$ and $\rho_n$ are specified by the two-point correlation function $\mathcal{C}(\mathbf{r}, \mathbf{r}^\prime; E_n)$. 

The form \eqref{eq:jointprobdistr} readily allows to calculate the probability of having the two PWAs of the $n^{\rm th}$ resonance with the same sign~\cite{jalabert14}, i.e., 
\begin{equation}
\label{eq:pospar}
\mathcal{P}(\gamma_n^{\mathrm{l}} \gamma_n^{\mathrm{r}} > 0) =
\frac{1}{2} + \frac{1}{\pi} \arcsin{(\rho_n)} \ .
\end{equation} 
Completely correlated [anti-correlated] PWAs corresponding to 
$\rho_n = 1$ [$\rho_n = -1$] lead to 
$\mathcal{P}(\gamma_n^{\mathrm{l}} \gamma_n^{\mathrm{r}} > 0) = 1$ 
[$\mathcal{P}(\gamma_n^{\mathrm{l}} \gamma_n^{\mathrm{r}} > 0) = 0$] 
and 
then, the condition ${\cal P}(D_n<0)=0$ for perfect sequences of phase-slips is verified.
In the uncorrelated case where $\rho_n = 0$, we have 
$\mathcal{P}(\gamma_n^{\mathrm{l}}\gamma_n^{\mathrm{r}} > 0) = 1/2$, so that
${\cal P}(D_n<0)=1/2$.

Assuming that $\rho_k$ is a smooth function of $k$ on the scale of
the mean wave-number difference between two resonances $\Delta k\approx \Delta E M/\hbar^2k \simeq \pi/kL^2$ (where we have taken 
$\mathcal{A}=L^2$), from Eq.\ \eqref{eq:probDn} we have \cite{jalabert14}
\begin{equation}
\label{eq:probDn2}
{\cal P}(D_k<0) \simeq 2 f(k) + \Delta k f'(k)
\end{equation} 
with
\begin{equation}
\label{eq:functf}
f(k) =
\frac{1}{4}-\frac{1}{\pi^2}\arcsin^2{\big(\rho_k\big)}\ .
\end{equation}
Since the extreme values of $\rho_k$ are $\pm 1$, the derivative
$\mathrm{d} \rho_k/\mathrm{d} k = 0$ there, such that Eq.\ \eqref{eq:probDn2} 
is well defined for all values of $\rho_k$.

\section{Partial-width amplitude correlator}
\label{sec:bc_pwa_correlator}

Adopting the Gaussian hypothesis discussed in the previous section, the parameters $\sigma_k^2$ and $\rho_k$ can be obtained from the kernel function ${\cal K}(\mathbf{r}, \mathbf{r}^\prime; \varepsilon)$ governing the probability density   
\eqref{eq:probdistrwf}, or directly from the two-point correlation function
${\cal C}(\mathbf{r}, \mathbf{r}^\prime; \varepsilon)$. Assuming that in the classically chaotic case the Wigner function is ergodically distributed on the allowed energy shell, the Voros-Berry conjecture \cite{voros76,berry77,toscano} provides an alternative to the maximum entropy approach to obtain Eq.~\eqref{eq:probdistrwf}, which in addition results in the prescription
\begin{equation}
\label{eq:correlator}
{\cal C}(\mathbf{r}, \mathbf{r}^\prime; \varepsilon) =
\frac{1}{\mathcal{A}} \  \ J_0(k|\mathbf{r} - \mathbf{r}^\prime|) 
\end{equation}
for the case of two-dimensional billiards. Here, $J_0$ is the zeroth Bessel function of the first kind. 

The assumption of a sharp localization of the Wigner function on the energy manifold has been shown to be a questionable approximation \cite{balazs,ozorio}. While extensive numerical work has verified the applicability of the resulting universal correlator for points $\mathbf{r}$ and $\mathbf{r}^\prime$ far away from the boundary \cite{aurich93,li94,srednicki96,backer98,backer02,backer08}, Eq.~\eqref{eq:correlator} fails to ensure the correct wave-function normalization \cite{gornyi02} or describing the situation where $\mathbf{r}$ or $\mathbf{r}^\prime$ approach the border of the billiard \cite{hortikar98,urbina04,urbina13}.

Using Eq.\ \eqref{eq:correlator} as a first approach to the problem allows to obtain simple expressions for $\sigma_k^2$ and $\rho_k$ in different regimes of the parameter $kL$ \cite{jalabert14}. The resulting $\sigma_k^2$ is a monotonic function of $k$, in sharp 
contrast with the numerical results of Fig.~\ref{fig:correlator_rho_glr}(a). The correlation $\rho_k$ 
obtained by this procedure exhibits
an oscillatory behavior as a function of $k$ on the scale of $2\pi/L$, with a growing amplitude. In the semiclassical regime of $kL \gg 1$, these oscillations are between $-1$ and $+1$, since $\rho_k \simeq \cos{(kL)}$. The resulting $kL$-dependence of the probability $\mathcal{P}(D_k<0)$ has the same tendency as the numerical results, but it exhibits a slower secular decrease with $kL$ \cite{jalabert14}. These limitations stem from the fact that Eq.~\eqref{eq:correlator} ignores the boundary effects, while the PWA correlator is determined by the wave-function correlator evaluated precisely at the entrance and exit of the QD. It is therefore important to go beyond the approximation used in Ref.\ \cite{jalabert14} and to provide an accurate estimation of $\rho_k$.
This is the goal pursued in this section.

For classically chaotic systems, in the small-$\hbar$ limit, the boundary corrections to Eq.\ \eqref{eq:correlator} can be addressed by writing the correlation function as \cite{hortikar98}
\begin{equation}
\label{eq:gsc}
\mathcal{C}(\mathbf{r}, \mathbf{r}^\prime; \varepsilon) = 
\frac{1}{\pi \ \varrho_{\rm sc}(\varepsilon)} \ 
{\rm Im} \ G_{\rm sc}(\mathbf{r}, \mathbf{r}^\prime; \varepsilon) 
\end{equation}
in terms of the smooth density of states $\varrho_{\rm sc}$ and the small-$\hbar$ limit $G_{\rm sc}$ of the Green function related by Eq.\ \eqref{eq:dos}.
In a semiclassical approach, $G_{\rm sc}$ can be split into its smooth component corresponding to the Green function of free space (direct path) and the contribution arising from the classical trajectories that experience the boundary \cite{hortikar98,urbina04,urbina13}. The first component results in the universal form \eqref{eq:correlator} of the wave-function correlator, to which we have to add a correction depending on classical trajectories different from the direct one~\cite{footnote1}. Such an approach has been used to address the problem of the peak-height modulation \cite{narimanov}, yielding oscillations related with the shortest classical trajectories joining the entrance and exit of the QD.

\begin{figure}[tb]
\centerline{\includegraphics[width=0.6\linewidth]{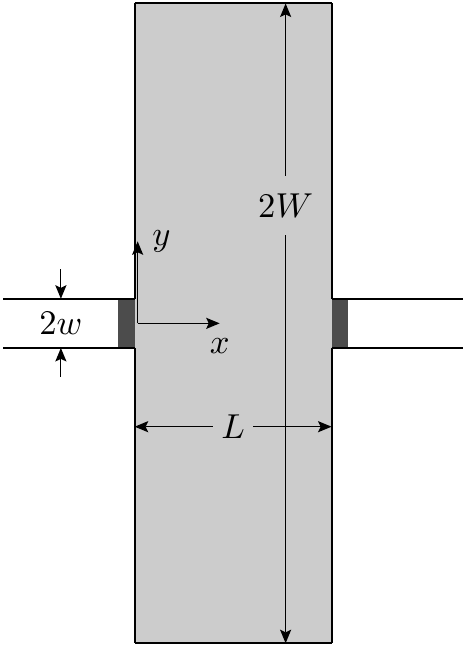}}
\caption{\label{fig:rectangle} Sketch of a
  rectangular quantum dot (light gray) of length $L$ and width $2W$, connected to leads of width $2w$ through tunnel barriers
  (dark gray).}
\end{figure}

In order to avoid the complicated implementation of semiclassical sums, we focus ourselves on an efficient description of the wave-function at the 
entrance and exit of the QD, and simply trade the geometry of the cavity in  Fig.\ \ref{fig:setup} by the rectangular billiard of width $2W$ depicted in Fig.\ \ref{fig:rectangle}. Taking the limit $W \rightarrow \infty$ eliminates the contributions from paths other than those going between the entrance and exit of the QD. We adopt the Dirichlet boundary condition for the horizontal walls and the Neumann one for the vertical walls, so that we are able to work with a separable system. The latter condition is the appropriate one for $|y| < w$, but not for $|y| > w$. When $W \rightarrow \infty$ the trajectories that ``see" the horizontal walls and/or the part of the vertical walls not connected to the leads become irrelevant, and despite our drastic simplification we expect that the most salient features of the wave-function correlation are well-described, since the trajectories between entrance and exit are the same for both geometries. 
In the Appendix we treat the case of finite $W$, where the integrability of the system drastically influences the probability of not observing phase-lapses between resonances. 

Assuming hard walls on the leads, the first transverse channel wave-functions read
\begin{equation}
\Phi_{\mathrm{l}}(y) = \Phi_{\mathrm{r}}(y) = \Phi_{1}(y) =
\frac{1}{\sqrt{w}} \ \sin{\left(Q_{1}[y-w]\right)} \, ,
\end{equation}
where the $a$\textsuperscript{th} transverse momentum in the leads is defined by $Q_a=\pi a/2w$. 
The eigenfunctions in the rectangle with mixed boundary conditions can be written as
\begin{equation}
\label{eq:psi_mb}
\psi_{m,b}(x,y) = \sqrt{\frac{2}{L}} \
\cos{\left(\frac{m \pi x}{L}\right)} \ \phi_{b}(y) \, ,
\end{equation}
with
\begin{equation}
\label{eq:phi_b}
\phi_{b}(y) = 
\frac{(-1)^p}{\sqrt{W}} \
\sin{\left(q_b[y-W]\right)} \, ,
\end{equation}
where $b=2p-1$ for odd $b$ and $b=2p$ for even $b$. The transverse wave-vectors in the rectangular cavity are defined by $q_b=\pi b/2W$ and the resonant energies are 
$\varepsilon_{m,b}=(\hbar^2 /2M)[(\pi m/L)^2+q_b^2]$.

In this simple geometry we can write the exact Green function. Using the spectral decomposition of the latter in the expression \eqref{eq:gsc} of the two-point correlation function, we have
%
\begin{equation}\label{eq:rhosq}
\rho_k = 
\frac{\langle \gamma_{\varepsilon}^{\mathrm{l}} 
\gamma_{\varepsilon}^{\mathrm{r}} \rangle}
{\langle \gamma_{\varepsilon}^{\mathrm{l}} 
\gamma_{\varepsilon}^{\mathrm{l}} \rangle} = 
\frac{\sum_{m,b} (-1)^m \delta(\varepsilon-\varepsilon_{mb}) \ \coola^2_{1b}}
{\sum_{m,b} \delta(\varepsilon-\varepsilon_{mb}) \ \coola^2_{1b}}
\, ,
\end{equation}
where we have defined the overlap of the transverse wave-functions
\begin{equation}\label{coola}
\coola_{1b} = \int_{-w}^{+w}\dif y \ \Phi_{1}(y) \ \phi_b(y)
= - \ \frac{2 Q_1 \ \cos{(q_b w)} }{\sqrt{wW} \ (q_b^2-Q^2_1)} \, .
\end{equation}
By symmetry, $\coola_{1b} = 0$ when $b$ is even. 

The overlap \eqref{coola} is characteristic of quantum problems with interfaces between two regions like the case of the conductance quantization through an abrupt quantum point contact~\cite{szafer89,gorini2013}. The overlap $\coola_{1b}$ presents, as a function of $q_b$, a maximum for $q_b \simeq Q_1$, and it is appreciably 
different from zero only in the interval $[0,Q_{2}]$. In the context of the conductance quantization, the previous observation has been used to justify the so-called mean field approximation (MFA) \cite{szafer89}, where 
\begin{equation}
\label{MFA1}
\coola_{1b}^{\mathrm{MFA}}
= \sqrt{\frac{w}{W}} \ \Theta(Q_2-q_b) \, .
\end{equation}
In the limit $W\to \infty$, the sums over $b$ in Eq.\ \eqref{eq:rhosq} can be readily performed by going to the continuum limit in $q_b$  while respecting the condition that $b$ must be odd. We then have 
\begin{equation}
\label{eq:rhoMFA}
\rho^{\mathrm{MFA}}_k = \frac{\displaystyle\sum_{m=m_\mathrm{min}}^{m_\mathrm{max}} \
\frac{(-1)^m}{\sqrt{k^2-(\pi m/L)^2} }}
{\displaystyle\sum_{m=m_\mathrm{min}}^{m_\mathrm{max}} \
\frac{1}{\sqrt{k^2-(\pi m/L)^2} }}
 \, ,
\end{equation}
with $m_\mathrm{min}=\left\lfloor(kL/\pi)\sqrt{1-(\pi/kw)^2}\right\rfloor+1$ and $m_\mathrm{max}=\left\lfloor kL/\pi\right\rfloor$, 
where $\lfloor \ldots\rfloor$ denotes the floor function. The resulting $\rho_k$ is represented by the brown line in Fig.~\ref{fig:correlator_rho_glr}(c). It provides a relatively good approximation to the value of $\rho_k$ obtained by fitting the local histograms of $z_n$ to Eq.\ \eqref{eq:probdistrproduct} calculated for the geometry of Fig.\ \ref{fig:setup} (green line). In the semiclassical limit of large $kL$, less and less values of $m$ fall into the allowed interval and the MFA becomes unreliable. This limitation can be overcome by tackling the evaluation of \eqref{eq:rhosq} without the simplification of the MFA. Treating again the sums over $q_b$ as continuous integrals we have
\begin{equation}
\label{eq:rhofull}
\rho_k = \frac{\displaystyle\sum_{m=0}^{m_\mathrm{max}} \
\frac{(-1)^m}{\sqrt{k^2-(\pi m/L)^2} } \
\frac{\cos^2{\left(w \sqrt{k^2-(\pi m/L)^2}\right)}}
{\left[ k^2-(\pi m/L)^2-(\pi /2w)^2 \right]^2}
}{\displaystyle\sum_{m=0}^{m_\mathrm{max}} \
\frac{1}{\sqrt{k^2-(\pi m/L)^2} } \ 
\frac{\cos^2{\left(w \sqrt{k^2-(\pi m/L)^2}\right)}}
{\left[ k^2-(\pi m/L)^2-(\pi /2w)^2 \right]^2}}
 \, .
\end{equation}
The resulting $\rho_k$ is represented by the red line in Fig.\ \ref{fig:correlator_rho_glr}(c) and provides an improved, and quite accurate, description of the numerically extracted $\rho_k$ (green line) for the geometry of interest. 

\begin{figure}[tb]
\centerline{\includegraphics[width=\linewidth]{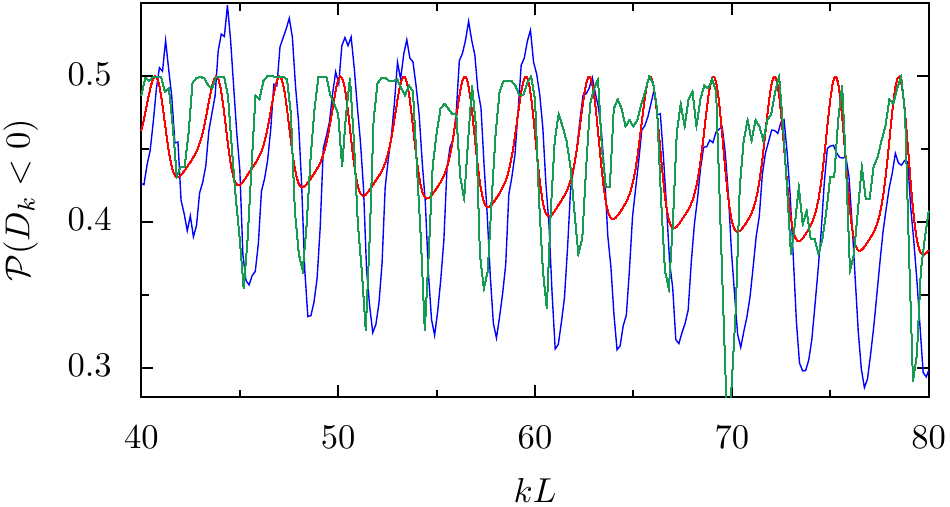}}
\caption{\label{fig:PDk_analytic_fitted} 
Probability $\mathcal{P}(D_k<0)$ of having a negative $D_k$ as a function of $kL$. 
The blue line reproduces the numerical data of Fig.\ \ref{fig:correlation}.
The green line uses $\rho_k$ from Fig.\ \ref{fig:correlator_rho_glr}(c) and Eq.\ \eqref{eq:probDn2}.
The red line uses $\rho_k$ from Eq.\ \eqref{eq:rhofull}, together with Eqs.\ \eqref{eq:probDn2} and \eqref{eq:functf}.}
\end{figure}

In Fig.\ \ref{fig:PDk_analytic_fitted} we compare the probability $\mathcal{P}(D_k<0)$ obtained in Sec.\ \ref{sec:transmissionphase} (see Fig.\ \ref{fig:correlation}) from the numerical data (blue line) with that resulting from the use of Eq.\ \eqref{eq:probDn2} together with an 
estimation of $\rho_k$. The green line is the result of using the estimation of Fig.\ \ref{fig:correlator_rho_glr}(c) for $\rho_k$, that is the correlator 
yielded by the fitting to the $k$-dependent histograms. The red line is the result of using Eq.\ \eqref{eq:rhofull} for estimating $\rho_k$. 
There is an overall agreement in the behavior of these curves, although not a detailed one. This result could be related to the poorer
statistics used in the $k$-dependent histograms, when compared to that of the numerical determination of $\mathcal{P}(D_k<0)$, as discussed in 
Sec.~\ref{sec:pwa_numerics}. Nevertheless, the use of any of the two estimations of $\rho_k$ results in a much closer approach to the numerical data, 
when compared with the case of Ref.\ \cite{jalabert14} in which $\rho_k$ is obtained by ignoring the boundary corrections to the wave-function correlator. 
The considerable improvement obtained with respect to the results of Ref.\ \cite{jalabert14} demonstrates the importance of such corrections.

\section{Conclusions}
\label{sec:ccl}

In this work we have established a link between two emblematic problems of quantum transport: the gate-voltage modulation of the peak-height and the phase-locking of the conductance between successive Coulomb blockade peaks. The failure to observe the universal behavior for the latter is correlated with small values of the average peak-height. Such
a result is restricted to the single-channel regime, characteristic of the Coulomb blockade physics. The established
connection stems from the dependence of both phenomena on the partial-width amplitude representing the coupling of the quantum dot to the leads, and it has been verified by a thorough numerical analysis. 

Having identified the PWAs as the key elements of the previous link, we numerically studied their statistical distribution, and determined that a joint probability density of the left and right PWA corresponding to a given resonance is relatively well described by a Gaussian law. Such a distribution is characterized by 
two parameters that depend on the resonance energy: the local average total width and the correlation between the two PWAs. 
The evaluation of the latter within a simple model provides a result in a good agreement with the numerically-extracted values. 

On the one hand, our work thus contributes to the important problem of wave-function correlations close to the boundary of a billiard \cite{urbina13}. On the other hand, it is experimentally relevant given the departures from a universal phase-locking of the conductance recently measured \cite{takada17,edlbauer2017} and in view of current experiments measuring the transmission phase in microwave cavities \cite{a_richter}.

\begin{acknowledgments}
We thank S.\ Takada, H.\ Edlbauer, and C.\ B\"auerle for useful discussions and communication of experimental data prior to publication. 
We thank J.-D.\ Urbina for discussions at the beginning of this work and for suggesting us to consider the geometry of Fig.\ \ref{fig:rectangle}. 
We acknowledge financial support from the ANR through grants ANR-11-LABX-0058\_NIE (Labex NIE) within ANR-10-IDEX-0002-02 and ANR-14-CE36-0007-01 (SGM-Bal), from the French-Argentinian collaborative project CNRS PICS 06687, 
and from the Spanish grant MINECO/FEDER No.\ FIS2015-63770-P.
\end{acknowledgments}

\setcounter{equation}{0}
\renewcommand{\theequation}{A\arabic{equation}}
\section*{APPENDIX: BEYOND THE CHAOTIC CASE}

The analysis presented in the main text has been based on the assumption that the underlying classical dynamics is chaotic, or at least sufficiently far away from regular. While such an hypothesis, based on the irregular shape of the working geometries and the influence of weak smooth disorder, has been validated by many experiments in ballistic transport \cite{scholarpedia}, it is interesting to discuss what is the behavior of the transmission phase in two extreme cases: that of a disordered QD and that of a regular billiard with integrable dynamics.

In the case of a disordered QD with short-range scatterers, characterized by an elastic mean-free path $\ell \ll L$, no phase-locking of the conductance 
is expected \cite{levy00}. Such a behavior can be understood from the fact that for a disordered system, the right-hand side of the wave-function correlator \eqref{eq:correlator} should be multiplied by the exponentially small factor $\mathrm{e}^{-\ell/L}$ \cite{mirlin00}, so that $\mathcal{P}(D_n<0) = 1/2$, 
implying that the presence and absence of phase-slips are equally likely \cite{molina12}.

As a representative case of the integrable dynamics, we now consider the rectangular cavity of Fig.\ \ref{fig:rectangle}, but in contrast to the analysis of Sec.\ \ref{sec:bc_pwa_correlator}, we do not take the limit of $W \rightarrow \infty$. Fundamental differences appear with respect to the case treated beforehand.

Integrable systems that are also separable have a simple nodal pattern and the condition \eqref{eq:condition} is straightforward to implement. Such is the case 
of the rectangular billiard with point contacts ($w\to 0$) where the PWA is simply proportional to the corresponding value of the resonance wave-function. Taking 
into account a finite width $2w$ of the leads, the system is no longer separable and the nodal structure can be very complex~\cite{albeverio96}.
In this Appendix we first consider the case of point contacts and then extend the results to the case of a finite lead width.

The closed rectangular billiard with Dirichlet conditions along the whole boundary has the same eigenenergies than those of Sec.\ \ref{sec:bc_pwa_correlator}.
Notwithstanding, instead of Eq.\ \eqref{eq:psi_mb}, the eigenfunctions take the form
\begin{equation}
\psi_{m,b}(x,y) = \sqrt{\frac 2L}\sin{\left(\frac{m\pi x}{L}\right)}\ \phi_b(y)\ ,
\end{equation}
with the transverse wave-function $\phi_b$ given by Eq.\ \eqref{eq:phi_b}.

\begin{figure}[tb]
 \centerline{\includegraphics[width=\linewidth]{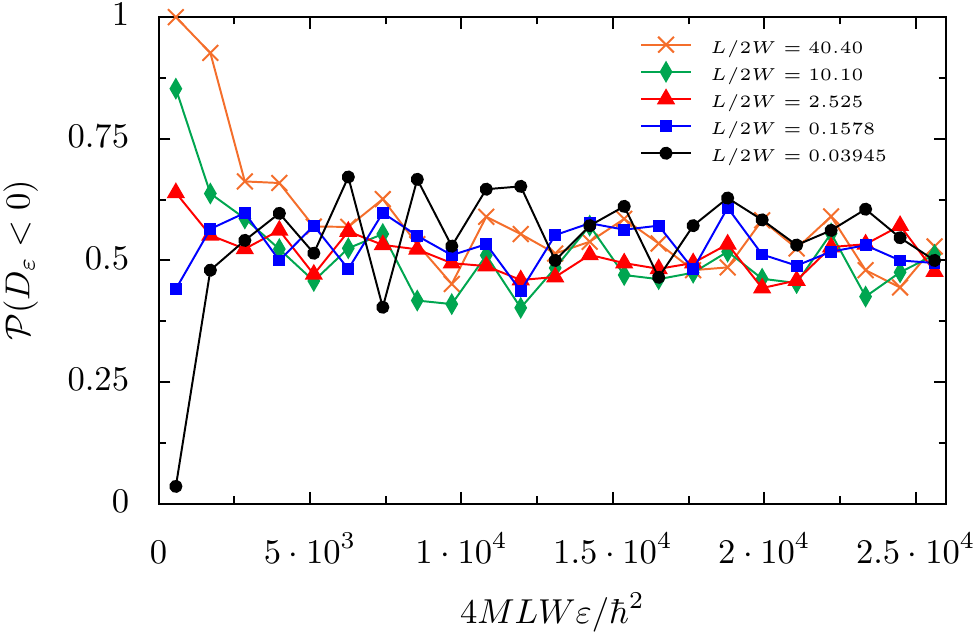}}
\caption{\label{fig:Prect} Probability $\mathcal{P}$ of not having a zero in between two resonances
for different values of $W$ and $L$, keeping a constant billiard area $\mathcal{A}=2WL$ so that
the eigenvalue densities are comparable. The leads are connected face-to-face 
on the two lateral sides.
The data points are calculated as one minus the number of zeros divided by the
number of resonances in equal energy intervals.
The asymptotic behavior for high energies is $\mathcal{P}=1/2$. For low energy, the first data point is close to
$\mathcal{P}=L/(2W+L)$ as argued in the text.}
\end{figure}

The corresponding nodal structure in this case is a checkerboard pattern with $m-1$ straight nodal lines in the $x$ direction and $b-1$ in the $y$ direction. 
Using the index $n$ to order the eigenstates by their eigenenergy, it
is easy to calculate whether $D_n$ is larger or smaller than $0$. With left and right leads connected to opposite walls in the $x$ direction of the rectangle,
we have
\begin{equation}
\label{eq:sign}
\mathrm{sgn}\{D_n\}=(-1)^{m_n+m_{n+1}}\ ,
\end{equation}
where $m_n$ denotes the number of nodes of the wave function $\psi_{m,b}$ in the $x$ direction.

Taking into account the discussion above, the geometry of the connection to the leads and the ratio between the side lengths
$2W$ and $L$ fully determine the probability $\mathcal{P}$ of not having a zero between two neighboring resonances. We assume incommensurate
values for $4W^2$ and $L^2$ to avoid degeneracies that complicate the situation. 

In the case of leads connected to opposite sides
of the rectangle separated by a distance $L$ one could naively think that, on average, for changing the number of nodes in the $x$
direction by one we need to increase the value of $b$ by $\Delta b=2W/L$. Hence, the probability of not having a zero between two resonances
could be written as $\mathcal{P}=L/(2W+L)$. However, this is only true at low energy. For higher energy values, 
very different combinations of $b$ and $m$ leading to quasi-degeneracies exist, so that the probability of having neighboring resonances
with zeros in between is close to $\mathcal{P}=1/2$. 

The combinatorial problem arises from the pecularities exhibited by the spectrum of the rectangle and is reminiscent of the spectral statistics of
the many-body non-interacting systems analyzed in Ref.\ \cite{munoz06}. In Fig.\ \ref{fig:Prect} we show numerical results for $\mathcal{P}$ 
for rectangular dots with different aspect ratios coupled to the leads at opposite sites of the transverse edges separated by a distance $L$. The geometry 
under study is the same as in Fig.\ \ref{fig:rectangle} of the main text. In the figure, we plot results as a function of the energy $\varepsilon$ instead of $kL$, 
since the values of $L$ are quite different for each of the dots considered. Each point is an average of the results for an energy window in a given rectangular billiard. It can be seen from Fig.\ \ref{fig:Prect} that, as the energy increases, the crossover between the low-energy case, depending on the aspect ratio of the billiard, and the asympotic limit where $\mathcal{P}=1/2$, is quite fast. The situation in regular, separable dots is then very different from the situation in chaotic QDs.

\begin{figure}[tb]
\includegraphics[width=\columnwidth]{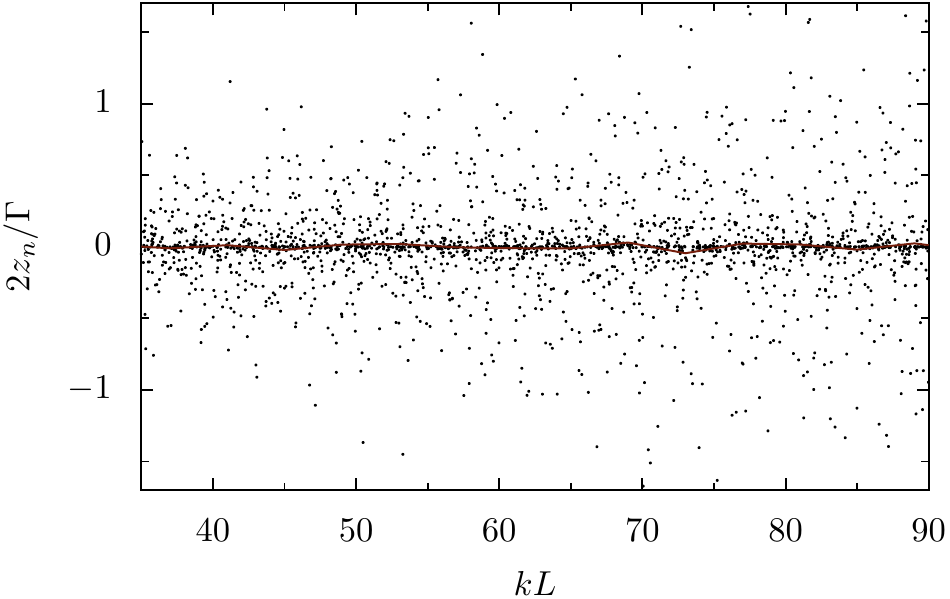}
\caption{\label{fig:regular} Individual values of the product $z_n=\gamma_n^l\gamma_n^r$ (black dots) normalized, as in Fig.\ \ref{fig:correlator_rho_glr}(b), by using one-half of the global average $\Gamma$ of the total width for an ensemble of rectangular billiards. The local average values of $z_n$ are indicated by the brown solid line, and do not exhibit the $kL$ modulation found for chaotic geometries.}
\end{figure}

Since the previous combinatorial and nodal structure arguments do not hold for the case with a finite-width connection to the leads, we have 
numerically computed the $z_n$- and $D_n$-distributions in such a case. 
In order to improve the statistics we considered 
an ensemble of eight rectangular billiards with finite-width contacts and incommensurate sides, all of them having similar lengths and areas. 
In order to avoid symmetry considerations that would complicate the results, we 
chose different geometries of the connection to the leads, trying to avoid symmetries. The results show that $\mathcal{P} \simeq 1/2$ for all values of $kL$. 
In Fig.\ \ref{fig:regular} we show the individual values of $z_n$, normalized like in Fig.\ \ref{fig:correlator_rho_glr}(b) by $\Gamma/2$, where the global average $\Gamma$ of the total width is taken over the whole $k$-interval. Like in Fig.~\ref{fig:correlator_rho_glr}(b), we observe an increase of the typical values of $z_n$ when increasing $k$, due to the increase of the barrier transparency and the use of the global normalization factor $\Gamma/2$. Unlike the case of Fig.\ \ref{fig:correlator_rho_glr}(b), the results of Fig.\ \ref{fig:regular} do not show a periodic modulation in $kL$, illustrating the important differences between the chaotic and integrable cases. Moreover, the local average of $z_n$ (brown line in Fig.\ \ref{fig:regular}) is structureless and very small, in contrast to the values of $2\sigma_k^2\rho_k/\Gamma$ of the chaotic case [green line in Fig.\ \ref{fig:correlator_rho_glr}(b)]. 

It is well known that the pecularitites of different integrable systems make it difficult to extract general conclusions. However, to the presented case of the rectangular 
billiard, we can add the results obtained for a triangular billiard with leads connected in different positions (not shown), which also lead to 
$\mathcal{P}\simeq 1/2$, 
and thus, an equally likely probability of observing or not phase-lapses. From the connection between the absence of phase-lapses and the low average 
height of the conductance peaks established in Sec.\ \ref{sec:transmissionphase}, we conclude that the modulation in the height of the Coulomb blockade peaks 
is not expected for integrable QDs, nor for disordered systems.


\begin{thebibliography}{99}
\bibitem{footnote:nodal_lines}
Alternatively, Eq.\ \eqref{eq:condition} can be recast in terms of the number of nodal lines of the resonance wave-functions existing between the 
left and right leads \cite{lee99}. If the difference in the number of nodal lines between consecutive states crossing any arbitrary curve joining the left and right leads through the dot is even (in particular if the number of nodal lines is the same), then $D_n>0$ and there is a phase-lapse between the resonances. If the difference is odd, then $D_n<0$ and there is no phase-lapse. 
The knowledge of the statistics of nodal lines \cite{aranovitch07, samajdar14} billiards can in principle be used to estimate the typical length of a sequence of resonances exhibiting phase-slips.
\bibitem{footnote:poorer_stat}
Unlike the numerical calculations of Sec.\ \ref{sec:correlation_numerics}, we do not make here any a priori
assumption on the $k$-periodicity and thus we do not mix the data of successive $k$-intervals. As a consequence, 
the gathered statistics for each $k$-dependent histogram is poorer than in the previous case.
\bibitem{footnote1}
When one of the points, say $\mathbf{r}^\prime$, is close to the boundary, the two-point correlation function can be written as 
${\cal C}(\mathbf{r}, \mathbf{r}^\prime; \varepsilon) =
(1/\mathcal{A}) \ [ J_0(k|\mathbf{r} - \mathbf{r}^\prime|) \mp 
J_0(k|\mathbf{r} - \mathbf{r}^{\prime\prime}|)]$, for Dirichlet (Neumann) boundary conditions, and $\mathbf{r}^{\prime\prime}$ being the mirror image of $\mathbf{r}^\prime$ upon reflection with respect to the boundary, the latter considered as an infinite, straight hard wall \cite{ishio02,urbina13}.
\end{thebibliography}
\end{document}